\documentstyle[12pt]{article}
\begin{document}
\newcommand{\be}{\begin{eqnarray}}
\newcommand{\ee}{\end{eqnarray}}
\def\lo{\langle 0 |}
\def\ro{ | 0 \rangle }
\def\fc{ f_{\eta'}^{(c)} }
\def\gmf{\gamma _{5}}
\def\la{\langle }
\def\ra{ \rangle }
\def\el{ \langle \eta'| }
\def\er{ | \eta' \rangle}
\def\gmmu{\gamma _{\mu}}
\def\gmnu{\gamma_{\nu}}
\def\gmf{\gamma _{5}}
\def\atop{ \frac{ \alpha_{s}}{4 \pi} G_{\mu \nu}
 \tilde{G}_{\mu \nu} }
 \begin{flushright}
hep-ph/9706316 \\
NI 97033-NQF
\end{flushright}
\begin{center}
   {\large \bf The gluon/charm content of the $\eta'$ meson }\\[2mm]
   {\large \bf and instantons }\\[5mm]

  {\large E.V.Shuryak$^{1,3}$ and A.R.Zhitnitsky$^{2,3}$}\\[5mm]
   {\small \it  $^1$Department of Physics,
   SUNY at Stony Brook,\\ Stony Brook, NY11790, USA \\
    $^2$Physics and Astronomy Department,
        University of British Columbia \\
         Vancouver, BC V6T 1Z1, Canada \\  
        $^3$ Isaac Newton Institute For Mathematical Sciences\\
        20 Clarkson Road, Cambridge, CB3 0EH, U.K.\\}
\end{center}
\vskip 1cm
\abstract{\noindent \small
Motivated by recent CLEO measurements
 of the 
$B\rightarrow \eta' K$ decay,
we evaluate gluon/charm content of the  $\eta'$ meson using the 
interacting instanton liquid model of the QCD vacuum.
 Our main result is
$\lo g^3 f^{abc} G_{\mu \nu}^a \tilde{G}_{\nu \alpha}^b 
G_{\alpha \mu}^c \er=(2.3\div 3.3) GeV^2\times
\lo g^2  G_{\mu \nu}^a \tilde{G}_{\mu \nu}^a \er$.
It is very large  due to the strong field 
of small-size 
instantons. We show that it provides quantitative explanations
of   the   CLEO data on the 
$B\rightarrow \eta' K$ decay rate (as well as  inclusive
process $B\rightarrow \eta'+ X$),  
 via a virtual Cabbibo-unsuppressed decay into $\bar c c$ pair
which then becomes  $\eta'$. If so, a significant
 charm component should be present in other hadrons also.
 In particular,  we found a large 
contribution  of the  charmed quark
in the $polarised$ deep-inelastic scattering on a
proton.   }
\vskip 1cm
%  \newpage
%  \section{Motivation}

1.Instantons of a small size ($\rho\sim 1/3 fm$) are
known for long time to be a very important
component of the QCD vacuum
\cite{Shu_82a}. In general, their fields are responsible for a scale 1 GeV
 which restrict perturbative QCD from below, and effective hadronic
Lagrangians from above. Because of fermionic zero modes, they play especially
important role for light (u,d,s) quark physics
 (for recent review see \cite{SS_96}). 
It was nevertheless  believed that
they are irrelevant for charm-related physics: and indeed,
 the instanton-induced spin-dependent and independent
potentials between heavy quarks are small compared to standard
confining-plus-perturbative  one. However, as we show in this paper,
the situation is reversed for $virtual$ $\bar c c$ pairs: they can only appear due
to the strongest gluonic fluctuations in vacuum,
and those are instantons. (In fact, the gluonic fields
in the centre of relevant instantons is so large, that one may
 even question whether $gG/m_c^2$ is  a good expansion parameter.)  

    The way to see this is to look at the charm component in hadrons with
different quantum numbers.
The object of this paper, $\eta'$,
  is long known to play a very
special role in QCD: separated by a large gap from other pseudo-scalars
(the Weinberg's U(1) problem \cite{Witten,Ven})
 it serves as a screening mass for the
topological charge (see recent detailed discussion in \cite{SV_95}).
   Thus testing whether the
high dimension gluonic
 operator does or does not couple
strongly  to the $\eta'$ we are actually testing 
whether the strongest vacuum fluctuations
do or do not possess  
the topological charge. No effect of such magnitude
 should exist e.g. for vector
 mesons: and indeed,  the empirical
 Zweig rule is very strict in vector channels,
allowing only tiny flavor mixing.

2.Recently, CLEO collaboration has reported \cite{CLEO}
 measurements of inclusive and exclusive production of 
the $ \eta' $ in B-decays :
\be
\label{1}
Br( B \rightarrow \eta' + X \; ; 2.2 \; GeV < E_{\eta'} <
2.7 \; GeV ) = (7.5 \pm 1.5 \pm 1.1) \cdot 10^{-4} \; ,  
\ee
\be
\label{2}
Br(B \rightarrow  \eta'+ K ) = (7.8_{-2.2}^{+
2.7} \pm 1.0) \cdot 10^{-5} \; .
\ee
%Here the inclusive branching ratio contains the acceptance 
%cut intended to reduce a background 
%from events with charmed meson interactions in a final state. At 
%first
%sight, the above numbers might seem quite innocent. However, 
Simple 
estimates \cite{HZ} show that these data are in  severe 
contradiction 
with the standard mechanism,
 the $b$-quark decay into 
 light quarks, because 
%which could be naively suggested  
% keeping in mind the standard picture of   
% $ \eta' $ as  a SU(3) singlet
%meson made of the $u-$, $d-$ and $s-$quarks. In this picture 
%the $ B \rightarrow \eta' $  amplitude must be proportional to the 
Cabbibo  suppression
factor $V_{ub}$ leads to
%and, as a result, the standard mechanism of 
%the $ B \rightarrow \eta' $ transition yields 
numbers which are 
by two orders of magnitude  smaller than the data (both the 
inclusive and exclusive cases). %Thus, there must
%be something beyond this standard picture.
    Alternative mechanism, suggested %ing a unified description of both
%the inclusive and exclusive modes, was proposed in the recent 
in \cite{HZ} %. It was shown that at the quark level $ B 
%\rightarrow \eta' $ decays can be described by
is based on the Cabbibo favored
$ b \rightarrow c \bar{c} s $ process, followed by a transition
of virtual $ \bar{c} c $ into the $ \eta' $. The latter transition 
may be possible, provided there exist
large intrinsic charm component of the $ \eta' $. Its  
 quantitative measure can be expressed through the matrix element
\be
\label{3}
\lo \bar{c} \gmmu \gmf c | \eta'(q) \ra \equiv i \fc q_{\mu} \; . 
\ee
and one needs
$\fc \approx 140 \; MeV$  
in order to explain the CLEO data, see \cite{HZ}.
 This value is surprisingly large, being 
 only a few times smaller than the 
analogously normalised residue $ \lo \bar{c} \gmmu \gmf c 
| \eta_c (q) \ra = i f_{\eta_c} q_{\mu} $ with $ f_{\eta_c} 
\simeq 400 \; MeV $ known experimentally from the $ \eta_c 
\rightarrow \gamma \gamma $ decay.

3.Because the $c$-quark is heavy, it may
only exist in the $ \eta' $ in a virtual loop,% Such a loop 
%with the heavy $c$-quark 
and its contribution can be evaluated 
in terms of gluonic  fields.
%background field technique. Therefore, 
%matrix element (\ref{3}) may not vanish due to $ \bar{c} c 
%\leftrightarrow gluons $ transitions. 
% To translate the charm operator $\bar{c} \gmmu \gmf c$ into the
% gluon content, it is convenient to start with
 Taking the divergence of the axial current 
in Eq.(\ref{3}) one gets
\be
\label{4}
\fc  = \frac{1}{m_{\eta'}^2} \lo 2 m_c \bar{c} i \gmf c +
\atop \er \; . 
\ee
% Since the $c$-quark is heavy, one can
which can be further simplified by the Operator Product
Expansion in inverse powers of the $c-$quark mass
\be
\label{5}
2 m_c \bar{c} i \gmf c = - \atop - \frac{1}{16 \pi^2
 m_{c}^2 }
g^3 f^{abc} G_{\mu \nu}^a \tilde{G}_{\nu \alpha}^b 
G_{\alpha \mu}^c 
+ O(G^4/m_{c}^4)  
\ee
(see the appendix in \cite{HZ} for a detailed derivation 
of this result. Further terms in expansion (\ref{5}) are 
neglected in what follows.)
Thus the problem is reduced to the 
matrix element of a particular dimension-6 pseudo-scalar gluonic operator:
 \be 
\label{6}
\fc = - \frac{1}{16 \pi^2 m_{\eta'}^2 } \frac{1}{m_{c}^2}
\lo g^3 f^{abc} G_{\mu \nu}^a \tilde{G}_{\nu \alpha}^b 
G_{\alpha \mu}^c \er \; . 
\ee

4.The  magnitude of the matrix
element (\ref{3})
 %can or cannot be explained by non-perturbative QCD. 
%attempted in  \cite{HZ}
%a line of arguments and approximations has related  the matrix element
%(\ref{7})
was  related  \cite{HZ} to the $vacuum$
 expectation value of similar  operators:
 \be
\label{8}
\fc \simeq \frac{3}{4 \pi^2 b} \frac{1}{m_c^2} \frac{
\la g^3 G^3 \ra _{YM}}{ \lo \atop \er }  \; . 
\ee
where
 $\la G^3\ra $ should be evaluated in pure gluodynamics, not QCD.
  Unfortunately, only indirect
 order-of-magnitude estimate for this latter quantity  was given, and thus
in \cite{HZ} rather
%leading to rather 
wide range of values  was given
$
\fc = ( 50 \div 180) \; MeV $.

5.We have performed direct calculation of this quantity
 using
the Interacting Instanton Liquid Model (IILM). In its present form,
 this model takes into account
 instantons coupling to light quarks to {\it all orders} in t'Hooft effective
 interaction,
which was shown to be crucial for $\eta'$ physics.
 It has correctly reproduced multiple mesonic/baryonic/glueball
 correlation functions,
 and also has an increasing direct support from
 lattice studies of instantons (see \cite{SS_96}).

 The calculation is based on numerical evaluation of the
following two-point Euclidean correlation functions
\be
\label{10}
K_{22}(x)=\lo g^2 G_{\mu\nu}^a \tilde {G_{\mu\nu}^a}(x)  g^2 G_{\mu\nu}^a \tilde {G_{\mu\nu}^a} (0)\ro 
\ee
\be
\label{11}
K_{23}(x)=\lo g^2 G_{\mu\nu}^a \tilde {G_{\mu\nu}^a}(x) g^3 f^{abc}G_{\mu\nu}^a \tilde{G}^b_{\nu\lambda}
G_{\lambda\mu}^c(0)\ro 
\ee
\be
\label{12}
K_{33}(x)=\lo g^3 f^{abc}G_{\mu\nu}^a \tilde{G}^b_{\nu\lambda}
G_{\lambda\mu}^c(x)~~ 
g^3  f^{abc}G_{\mu\nu}^a \tilde{G}^b_{\nu\lambda}
G_{\lambda\mu}^c (0)\ro
\ee
Studies of $K_{22}(x)$ has been made previously
 \cite{SS_95}, where it was demonstrated   that
in the ``unquenched'' ensemble of instantons with dynamical quarks
the non-perturbative part change sign at distances $x>0.6 fm$,  
displaying a ``Debye cloud'' of compensating topological charge.
It is identified with the $\eta'$ contribution, and lead to an estimate
\be 
\label{12a}
\lo g^2  G_{\mu \nu}^a \tilde{G}_{\mu \nu}^a \er =
\frac{16\pi^2}{\sqrt{3}}f_{\eta'} m_{\eta'}^2 \approx 7\, GeV^3
\ee  
 which
agrees reasonably well with other estimates in literature. 
 In this formula we have expressed matrix element (\ref{12a})  in terms of the  
  standard parameter $f_{\eta'}\approx 85 MeV$
which is defined as follows
$$
\lo \frac{1}{\sqrt{3}}\sum_{i=u,d,s} \bar{q}_i
 \gmmu \gmf q_{i} \er = i f_{\eta'} q_{\mu}.
$$
Using an anomaly in the chiral limit, $m_u=m_d=m_s=0$ we arrive to (\ref{12a}).

6.   We have calculated the correlators mentioned
  by numerical simulation, using of the ensemble 
16 instantons and 16 anti-instantons, put into a box $4\times 2^3 fm^4$,
 with  (without)  dynamical quarks\cite{c1}
Unfortunately, 
 the propagation
 of the  gluons in the background non-perturbative fields of instantons
  was not studied
in such details as for light quarks, and so far we do not  have
the gluon propagator program which could be used  for all distances.
%, and so
%one needs an extra care and approximations. Two  limits are simple.
At small x  purely perturbative results  (e.g.
 $K^{pert}_{22}(x)= 384 g^4/\pi^4 x^8$ ) dominate, while the
 non-perturbative  
fields can be included via
the operator product expansion (see e.g.\cite{NSVZ,Shu_82a}).
At large $x$ we would argue below that (at least with dynamical quarks)  
the non-perturbative fields dominate.

 The quantity  $\fc$ (\ref{6}) can be obtained from
%we arrive to the following expression for those matrix elements
%in terms of
 the correlation functions (\ref{10}, \ref{11},
\ref{12})\cite{c2}:
\be
\label{15}
\frac{\fc\sqrt{3}m_c^2}{f_{\eta'}}=\frac{K_{23}(x\rightarrow\infty)}
{K_{22}(x\rightarrow\infty)}=
%\frac{\fc\sqrt{3}m_c^2}{f_{\eta'}}=+
\sqrt{\frac{K_{33}(x\rightarrow\infty)}
{ K_{22}(x\rightarrow\infty)}}
\ee
   It is expected that at  large distances the contribution to 
  two other correlators would also be dominated by non-perturbative field
of the instantons. If so, one has a simple estimate for the ratio
of matrix elements
\be \label{est}
{\lo g^3 f^{abc} G_{\mu \nu}^a \tilde{G}_{\nu \alpha}^b 
G_{\alpha \mu}^c \er \over
\lo g^2  G_{\mu \nu}^a \tilde{G}_{\mu \nu}^a \er}={12\over 5} 
\la {1\over \rho^2} \ra \approx  (1\div 1.5) GeV^2
\ee
Two numbers given here correspond to
  averaging over instanton size distribution
for two variants of the instanton-anti-instanton
interaction, the so called ``streamline'' and ``ratio-ansatz'' ones, and
indicate the sytematics involved.
The latter 
(giving smaller average size and larger number above)
should be considered preferable, because it
 better agrees with the size distribution directly
obtained from lattice gauge field configurations, see  discussion in 
\cite{SS_96}. (Recent measurements \cite{DH_97}
 using refined ``inverse blocking'' method has found somewhat 
smaller instantons than others, but those seem to belong to correlated instanton-anti-instanton pairs, which would not contribute to the compensating Debye cloud we look for.)

  In our  measurements of $K_{23},K_{33}$
 $both$ ratios entering (\ref{15})  were found to
 stabilize at large enough x$>0.8 fm$ at the $same$ numerical value.
We take it as an indication that  $\eta'$  contribution
does in fact dominate, although 
 we were not able to see that all correlators fall off
 with the right mass\cite{c3}.
 Numerical values of the ratios
  about $(1.5\div2.2 )GeV^2$, for two ensembles mentioned.
The numers are somewhat larger than in (\ref{est})
because the second operator
in the correlator makes it more biased toward smaller instantons.

Proceeding to final result, we have to look at radiative corrections.
 The experimental number mentioned above is defined at
 the scale $\mu_1^2\approx m_c^2$, which is
 different from that obtained in the instanton calculation. In the latter case the charge and fields are normalized at
$\mu_2^2\approx gG$ where G is the typical gauge field at the points which
contribute the most to the correlators.
 Two scales are not too  far apart numerically 
 $\mu_2^2\approx (0.5\div 1 ) GeV^2 $, but the   anomalous dimension of the 
%%%%%$ G^3 $ operator as well as 
$g^3G\tilde{G}G$ operator  \cite{Mor} is large, and it leads to correction    
\be
\label{correction}
\fc (\mu_1\simeq m_c)=[\frac{\alpha_s(\mu_1)}{\alpha_s(\mu_2)}]^{\frac{-18}{2b}}
\fc (\mu_2)
\simeq 1.5\fc (\mu_2) ,
\ee
Here we use $m_c(\mu_1\simeq m_c)\simeq 1.25 GeV$ for the numerical estimates.
This concludes our derivation of the parameter
 \be
 \label{result}
\frac{\fc}{f_{\eta'}}\simeq 0.85\div 1.22 ,
\ee
where the second  value is preferable, see above. 
We present our final result  as the ratio $ \fc /f_{\eta'}$
instead of  the absolute value of $\fc$ because most systematic errors are gone
for the ratio.    
 	The  final uncertainty in eq. (\ref{result}) comes from the
systematic errors of the instanton model,
which can be judged from
comparison of the instanton size distribution or the scalar glueball size
to corresponding
lattice results. It will certainly be soon reduced by on-going works.
Finally, we compare it with the ``experimental'' value
needed to explain CLEO measurements (\ref{3}), and conclude that
our result
obtained in the instanton liquid model  
   agrees with it, inside the uncertainties.

7.
 The next logical question to ask is whether
the unexpectedly large gluon/charm content of $\eta'$
has profound consequences outside of $B$ physics,  for other hadrons. 
One point we want to make is that that it seems now likely that
 understanding of the spin  problem of the nucleon cannot be done without
its ``intrinsic charm'' as well\cite{3}. Relevant
 matrix element 
\be
\label{a1}
\la N  | \bar{c} \gmmu \gmf c | N \ra = 
g_A^{(c)}  \bar{N}      \gmmu \gmf N 
\ee
could be generated by the $ \eta' $ ``cloud'' inside the nucleon. 
%large. 
Assuming now the  $ \eta' $ dominance in this matrix 
element\cite{c4}
one could get the following 
   Goldberger-Treiman type relation\cite{3}
$
g_A^{(c)} = \frac{1}{2 M_{N}} g_{\eta' NN} \fc
$.
Although the precise value
of $ g_{\eta' NN } $ is unknown, and 
phenomenological estimates of the coupling   
% $ g_{\eta' NN } $
 vary significantly
$ g_{\eta' NN } = 3-7 $\cite{gNN}, it leads to    
%  Using numerical value 
%(\ref{20}), one could estimate $g_A^{(c)}$\cite{3}:
\be
\label{a3}
 \la N  | \bar{c} \gmmu \gmf c | N \ra = (0.2\div 0.5)
\bar{N}     \gmmu \gmf N   
\ee
%This result has a large uncertainty mainly due to a poor 
%knowledge of $ g_{\eta' NN} $. Nevertheless it shows
%that $g_A^{(c)}$ could be large and 
which is comparable to the light
 quark contribution! We plan to calculate  $g_A^{(c)}$ and $g_{\eta' NN }$
in the instanton model as well. Lattice determination of all those 
quantities would be more than welcome. Ultimately, 
the contribution of the charmed quarks 
in polarized deep-inelastic scattering may be  tested
experimentally, by tagging the charmed quark jets.

8.It is by now widely known that 
the Zweig rule is badly broken  in all 
scalar/pseudoscalar channels, and  that (rather large) mass of the $\eta'$
is in fact due to light-quark-gluon mixing. Furthermore,
all these phenomena are attributed to instantons.
In this work we have found that 
similar phenomena are even more profound for larger-dimension
 (multi-gluon) operators as well.
Moreover, the flavor mixing includes also a significant
 fraction of  $\bar c c$ in $\eta'$.
 Perhapse it is not so surprising  qualitatively:  but the fact
that one can actually   quantitatively calculate
these matrix elements  
and quantitatively compare it to real data is still 
rather amasing.

\section*{Acknowledgments}
This work was done during the program in I.Newton institute in
Cambridge,
and we use this opportunity to thank P.van Baal for invitation and
nice
program.This work is partly supported by the 
US Department
 of Energy under Grant No. DE-FG02-88ER40388.


\begin{thebibliography}{99}
\bibitem{c1}In order to avoid
$\eta-\eta'$ mixing, we use 3 flavors of quarks with the same mass.
\bibitem{c2} We should note
that the minus sign appears twice: in   definition $\fc$ (\ref{6})
and in the transition from Euclidean to Minkowski space. Therefore, 
sign has disappeared from the final expression (\ref{15}).
\bibitem{c3}These distances
are clearly large enough for the $\eta'$ intermediate state
 to be clearly separated from the
contribution of the pseudoscalar glueballs: as calculations without 
dynamical fermions shows, their mass is 2.4-3 GeV. The problem here is 
 technical:
it is difficult to get small signal out of statistical noise, and to 
correct properly for propagation in the finite volume.
\bibitem{c4}Note that such a saturation becomes exact 
in the 
large $ N_c $ limit.
\end{thebibliography}
\end{document}